\documentclass[aip,pop,amsmath,reprint]{revtex4-1}
\usepackage{graphicx}
\usepackage{epstopdf}
\usepackage{hyperref}
\usepackage[caption=false]{subfig} 

\newcommand{\eqr}[1]{Eq.\thinspace(#1)}
\newcommand{\refcite}[1]{Ref.\thinspace[\onlinecite{#1}]}

\hypersetup{colorlinks=true, allcolors=blue}

\begin{document}

\title{A Gyrokinetic 1D Scrape-Off Layer Model of an ELM Heat Pulse}

\author{E. L. Shi}%
\email{eshi@princeton.edu}%
\affiliation{%
\mbox{Department of Astrophysical Sciences, Princeton University, Princeton, NJ 08543, USA}}%

\author{A. H. Hakim}%
\author{G. W. Hammett}
\affiliation{Princeton Plasma Physics Laboratory, Princeton, NJ 08543, USA}
\affiliation{\mbox{Max-Planck/Princeton Center for Plasma Physics, Princeton
 University, Princeton, NJ 08543, USA}}

\date{\today}

\begin{abstract}
  An electrostatic gyrokinetic-based model is applied to simulate
  parallel plasma transport in the scrape-off layer to a
  divertor plate. The authors focus on a test problem that has been studied
  previously, using parameters chosen to model a heat
  pulse driven by an edge-localized  mode (ELM) in JET.
  Previous work has used direct particle-in-cell
  equations with full dynamics, or Vlasov or fluid equations
  with only parallel dynamics. With the use of the gyrokinetic
  quasineutrality equation and logical sheath
  boundary conditions, spatial and temporal resolution requirements
  are no longer set by the electron Debye length and plasma frequency,
  respectively. This test problem also helps illustrate some of the
  physics contained in the Hamiltonian form of the gyrokinetic
  equations and some of the numerical challenges in developing an edge
  gyrokinetic code.
\end{abstract}

\pacs{}

\maketitle

\section{Introduction}
One of the major issues for ITER and subsequent higher-power tokamaks
is the power load on plasma-facing components (PFCs) from energy
expelled into the scrape-off layer (SOL) by edge-localized modes.
Excessive total and peak power loads from ELM heat pulses can cause
the erosion or melting of divertor targets. Large Type I ELMs can
also result in erosion to the main chamber wall and the release of
impurities into the core plasma.\cite{Pitts2005} Suppressing ELMs
or mitigating the damage they cause to PFCs is crucial for the viability of
reactor-scale tokamaks. An accurate prediction of heat
fluxes on future devices is important for the development of
mitigation concepts.

Numerical simulations of heat pulse propagation can provide useful
information about the time dependence of the power load on divertor
targets. A test case involving the propagation of a heat pulse from
an ELM along a scrape-off layer to a divertor target plate has been
used as a benchmark in recent literature. This problem was first
studied using a particle-in-cell (PIC) code and was demonstrated to have good agreement
with experiment.\cite{Pitts2007} A Vlasov-Poisson model was later
developed to study this problem.\cite{Manfredi2011} A benchmark of
fluid, Vlasov, and PIC approaches to this problem was recently
described in \refcite{havlickova}.
An implementation of this test case in BOUT++ was used to compare non-local
and diffusive heat flux models for SOL modeling.\cite{Omotani2013}
With the exception of initial conditions, the parameters we have adopted for our simulations are
described in \refcite{havlickova}.
This test case involves just one spatial dimension (along the field line),
treating an ELM as an intense source near the midplane without trying to
directly calculate the magnetohydrodynamic instability and reconnection processes that
drive the ELM. Nevertheless, this is a useful problem for testing codes
and understanding some of the physics involved in parallel propagation
and divertor heat fluxes.

Unlike previous approaches, we have developed and studied
gyrokinetic-based models with sheath boundary conditions using
fully kinetic electrons or by assuming a Boltzmann response for the electrons.
As is often done in gyrokinetics (unless looking at very small electron-scale
turbulence where quasineutrality does not hold), a gyrokinetic quasineutrality equation
(which includes a polarization-shielding term) is used, so the Debye length does not need to be resolved.
To handle the sheath, logical sheath boundary conditions\cite{Parker1993} are used, which maintain zero net
current to the wall at each time step.  Although our simulations are
one-dimensional, perpendicular effects can be incorporated by assuming
axisymmetry. In an axisymmetric system, poloidal gradients have
components that are both parallel and perpendicular to the magnetic
field. The perpendicular ion polarization dynamics then enter the field equation by accounting for
the finite pitch of the magnetic field.

An advantage of the models we have developed is their low
computational cost. Earlier kinetic models have been described as
computationally intensive \cite{Pitts2007} due to restrictions in the time
step to $\sim \omega_{pe}^{-1}$ and in the spatial resolution to $\sim
\lambda_{De}$. (A 1D Vlasov model using an asymptotic-preserving
implicit numerical scheme described in \refcite{Manfredi2011} was able to
relax these restrictions somewhat for this problem, using
$\Delta x \sim 2 \lambda_{De}$ and $\Delta t \sim 4 / \omega_{pe}$ because
their simulation still included the sheath directly.)
By using a gyrokinetic-based model and logical sheath boundary
conditions, our code can use grid sizes and time steps that are several
orders of magnitude larger than this.
It is fully explicit at present, though one could consider extending
it to use implicit methods (such as in \refcite{Manfredi2011}) in the future.
While fluid models have their own merits,
they miss some kinetic effects, including the effect of hot tail electrons
on the heat flux on the divertor plate and the subsequent rise of sheath potential.

We have implemented our models in {\tt Gkeyll}, a code employing
discontinuous Galerkin (DG) methods that is being developed for several
applications, including solving gyrokinetic equations in the edge region.
Although {\tt Gkeyll} is currently being extended to have 5D capability,
we focus on $1x$+$1v$ $(v_{\parallel})$ simulations in this paper for comparison
with the similar $1x$+$1v$ Vlasov code in \refcite{havlickova}.

An explicit third-order strong-stability-preserving Runge-Kutta algorithm
is used to advance the system in time.\cite{Gottlieb2001} A review of the
Runge-Kutta DG algorithm is given by Cockburn and
Shu.\cite{Cockburn:2001vr} Our modifications to the basic DG scheme are
applicable to a general class of Hamiltonian evolution equations and
conserve energy exactly even when upwind fluxes are used (in addition
to conserving particles exactly).
These details will be described in a future publication.

Gyrokinetic codes that are fairly comprehensive (including general magnetic
fluctuations to varying degrees) have been developed\cite{Dorland2000,
 Kotschenreuther1995, Jenko2000, Candy2003,Candy2003-jcp, Parker2004,
 Peeters2009, Bottino2010, Maeyama2013} for the main core region of
fusion devices and have been fairly successful in explaining core turbulence in many
parameter regimes.
However, extensions are needed to handle the additional complexities of the edge
region ($r/a > 0.9$), such as open and closed field lines, plasma-wall
interactions, large amplitude fluctuations, and electromagnetic
fluctuations near the beta limit. The test problem studied here is a
useful first step in testing gyrokinetic algorithms for the edge
region. Such a code could also be used to simulate linear devices (such
as LAPD\cite{Umansky2011} and Vineta\cite{Kervalishvili:2006}) used
for studying fundamental plasma physics phenomena.

Section \ref{sec:esmodel} describes an electrostatic 1D gyrokinetic-based
model with a modification to the ion-polarization term to set a
minimum value for the wave number. Numerical implementation details
and the logical sheath boundary condition are described in Section
\ref{sec:numerical}. Results from numerical simulations and specific
initial conditions are presented in Section \ref{sec:results}.

\section{Electrostatic 1D gyrokinetic model with kinetic electrons\label{sec:esmodel}}
In this paper, we focus on the long-wavelength-drift-kinetic limit of gyrokinetics and ignore
finite-Larmor-radius effects for simplicity. Polarization effects are kept in the
gyrokinetic Poisson equation, and the model has the general form
of gyrokinetics and can be extended to include full gyroaveraging in the future.

\begin{figure*}
\centering
 \subfloat[]{\label{fig:geoA}\includegraphics[height=17em]{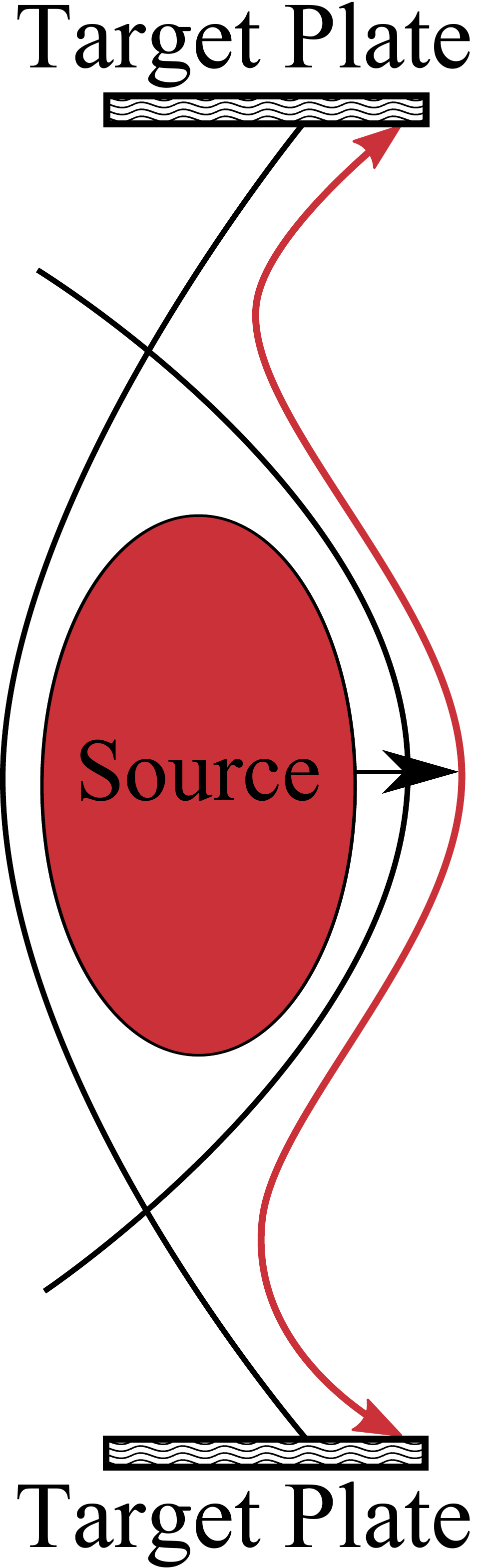}}
 \qquad
 \subfloat[]{\label{fig:geoB}\includegraphics[height=17em]{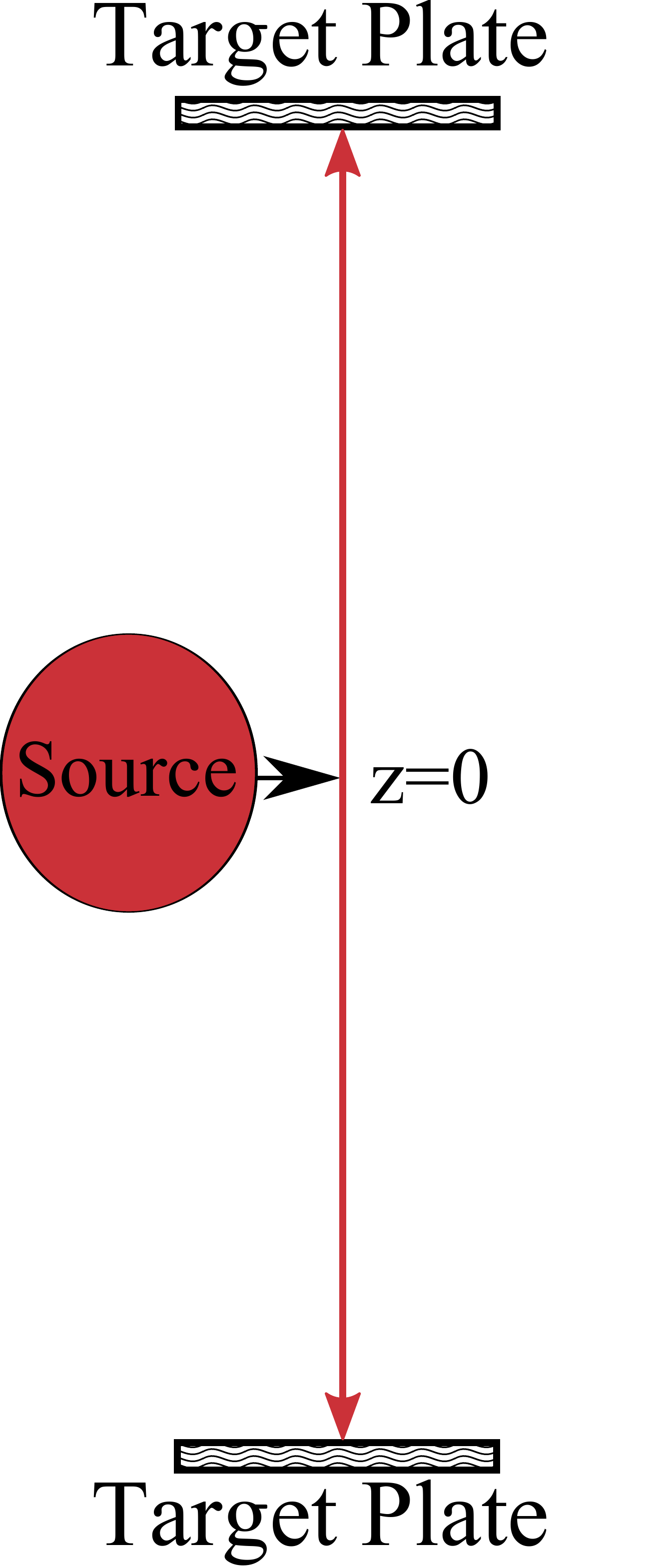}}
 \subfloat[]{\label{fig:geoC}\includegraphics[width=0.5\textwidth]{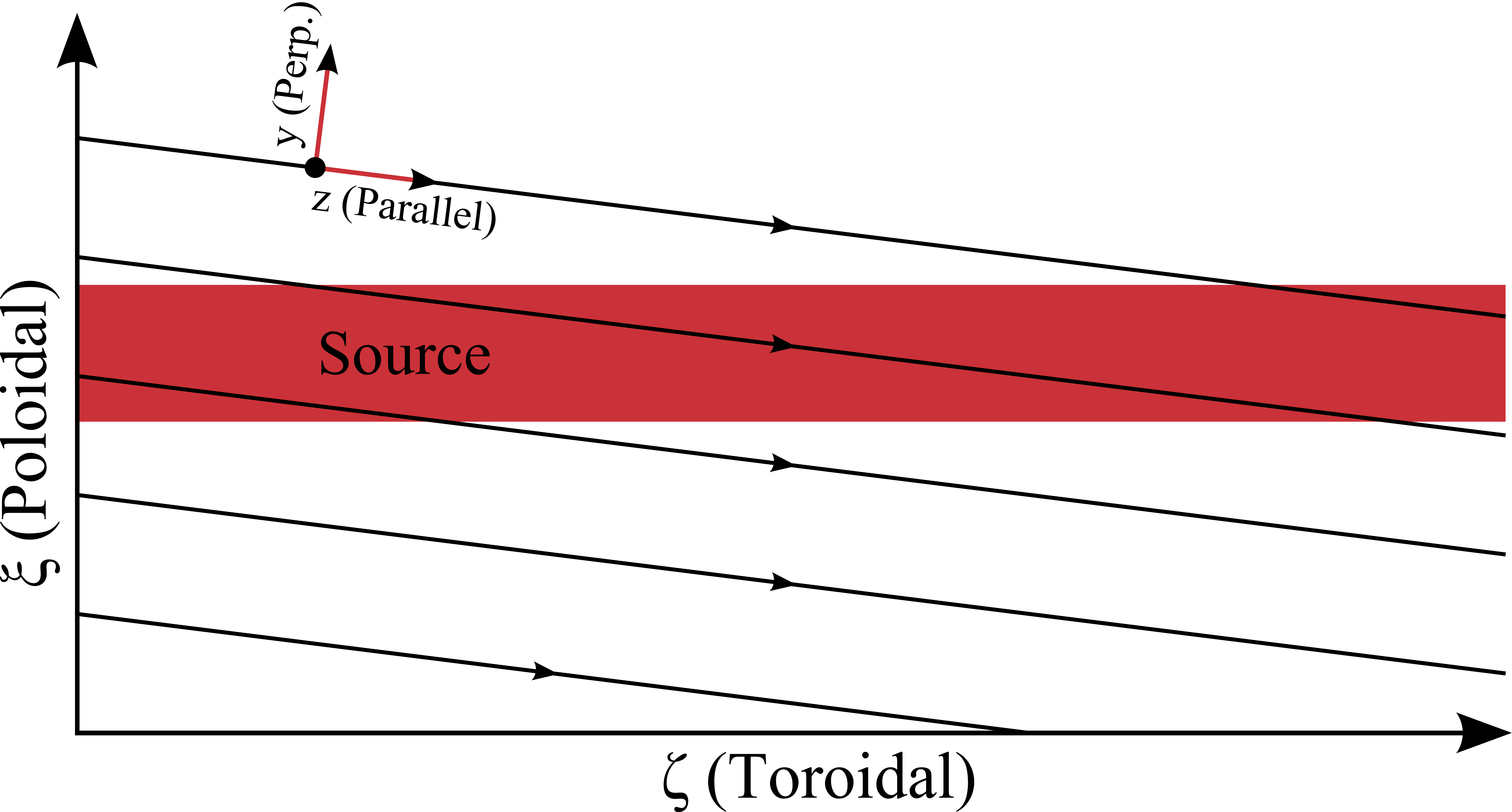}}
\caption{\label{fig:geo} Illustration of the geometry used in the ELM
 SOL heat pulse test problem. The scrape-off layer region in the
 poloidal cross section \protect\subref{fig:geoA} is treated as straight \protect\subref{fig:geoB} in this test,
 with the ELM represented by an intense source near
 the midplane region. The time history of the resulting heat flux to
 the target plate is calculated in the simulation. The side view \protect\subref{fig:geoC} illustrates
 that although there is no toroidal variation in this axisymmetric problem, poloidal variations
 lead to both parallel and perpendicular gradient components.}
\end{figure*}

The geometry used in the ELM SOL heat pulse test problem is illustrated in
Fig. \ref{fig:geo}. The Vlasov and fluid codes used in \refcite{havlickova}
consider only the parallel dynamics, while the 1$x$-3$v$ PIC code used in
\refcite{havlickova} includes full orbit (not gyro-averaged) particle dynamics
in an axiysmmetric system and so would automatically include
polarization effects on time scales longer than an ion gyroperiod.

The gyrokinetic equation can be written as a Hamiltonian evolution
equation for species $s$ of a plasma
\begin{align}
 \frac{\partial f_s}{\partial t} = \{H_s, f_s \} \label{eqn:gk},
\end{align}
where $H_s =
p_\parallel^2 / 2 m_s + q_s\phi - m_s V_E^2 / 2$ is the Hamiltonian
for the 1D electrostatic case considered here,
$p_\parallel = m_s v_\parallel$ is the parallel momentum, and
$\{f,g\} = (\partial f / \partial z) (\partial g / \partial p_\parallel) -
(\partial f / \partial p_\parallel) (\partial f / \partial z)$ is
the Poisson bracket operator for any two functions $f$ and $g$. The
potential is determined by a
gyrokinetic Poisson equation (in the long-wavelength quasineutral limit):
\begin{align}
-\partial_\perp \left( \epsilon_\perp \partial_\perp \phi \right)
 = \frac{\sigma_g}{\epsilon_0} = \frac{1}{\epsilon_0} \sum_s q_s \int d v_\parallel \, f_s \label{eqn:gk-Poisson}.
\end{align}

Here, $\sigma_g$ is the {\em guiding-center} charge density, while the
left-hand side is the negative of the polarization contribution to the
density, where the plasma perpendicular dielectric is
\begin{align}
\epsilon_\perp = \frac{c^2}{v_A^2} = \sum_s \frac{n_s m_s}{\epsilon_0 B^2}.
\label{eqn:epsilon_perp}
\end{align}
The ion polarization dominates this term, but a sum
over all species has been included for generality.

In the Hamiltonian, $V_E = - (1/B) \partial_\perp \phi$ is the $\boldsymbol{E}
\times \boldsymbol{B}$ drift in the radial direction (out of the plane in
Fig. \ref{fig:geoC}). Since there
is no variation in the radial direction, there is no explicit
$\boldsymbol{V}_E \cdot \nabla$ term, and $V_E$ only enters through the
second-order contribution to the Hamiltonian, $-m V_E^2 / 2$.
\refcite{Krommes13,Krommes12} provide some physical interpretations of
this term, and \refcite{Krommes13} gives a derivation of it in the cold-ion limit.

The conserved energy is given by
\begin{align}
W_{\rm tot} &= \int dz \sum_s \int dv_\parallel \, f_s H_s \nonumber\\
 &= W_K + \int dz \, \sigma_g \phi - \frac{1}{2}\int dz \, \rho V_E^2,
\end{align}
where $W_K = \int dz \sum_s\int dv_\parallel \, f_s m_s v_\parallel^2 /
2$ is the kinetic energy, and $\rho$ is the total mass density. Using the
gyrokinetic Poisson equation (\ref{eqn:gk-Poisson}) to substitute
for $\sigma_g$ in this equation and doing an integration by parts (with
a global neutrality condition $\int dz \sigma_g = 0$ so boundary terms
vanish), one finds that the total conserved energy can be written as
\begin{align}
W_{\rm tot} &= \frac{1}{2}\int dz \sum_s \int dv_\parallel \, f_s \left(m_s v_\parallel^2 + m_s V_E^2\right) \nonumber\\
 &= W_K + \frac{1}{2}\int dz \, \rho V_E^2.
\end{align}

To verify energy conservation, first note that $\int dz \int d
v_\parallel H_s \partial f_s / \partial t = 0$ by multiplying the gyrokinetic equation (\ref{eqn:gk}) by
the Hamiltonian and integrating over all of phase-space.
(Here, periodic boundary conditions are used for simplicity; there are of
course losses to the wall in a bounded system.)
The rate of change of the total conserved energy is then written as
\begin{align}
\frac{d W_{\rm tot}}{dt} &= \int dz \sum_s \int dv_\parallel \, f_s \left(
q_s \frac{\partial \phi}{\partial t} - \frac{m_s}{2} \frac{\partial V_E^2}{\partial t}
\right) \nonumber\\
&= \int dz \, \left( \sigma_g \frac{\partial \phi}{\partial t} 
 - \frac{1}{2} \sum_s n_s m_s \frac{\partial V_E^2}{\partial t}\right).
\end{align}
Using the gyrokinetic Poisson equation (\ref{eqn:gk-Poisson}) to substitute for $\sigma_g$
and integrating by parts, one finds that these two terms cancel, so $d W_{\rm tot}/dt = 0$.
Note that the small second-order Hamiltonian term $H_2 = -(m/2) V_E^2$
was needed to get exact energy conservation.
(In many circumstances, the $\boldsymbol{E}
\times \boldsymbol{B}$ energy $m V_E^2$ is only a very small
correction to the parallel kinetic energy $m v_\parallel^2/2$, but it is
still assuring to know that exact energy conservation is possible.)
This automatically occurs in the Lagrangian field theory approach to
full-$F$ gyrokinetics\cite{Sugama2000,Brizard_variational,Krommes12}, in which the
gyrokinetic Poisson equation results from a functional derivative of the action with
respect to the potential $\phi$, so a term that is linear in $\phi$ in
the gyrokinetic Poisson equation comes from a term that is quadratic in $\phi$ in the Hamiltonian.

\subsection{Electrostatic model with a modified ion polarization term}
One can obtain a wave dispersion relation by linearizing
Eqs.\thinspace({\ref{eqn:gk}}) and (\ref{eqn:gk-Poisson}) and Fourier
transforming in time and space.
With the additional assumption that $q_e = -q_i$ and neglecting ion
perturbations (except for the ion polarization density), one has
\begin{align}
k_\perp^2 \rho_s^2 + \left[1 + \xi Z\left(\xi\right)\right]&= 0.
\end{align}
Here, $\rho_s^2 = T_e/(m_i \Omega_{ci}^2)$, $\xi = \omega/(\sqrt{2}
k_\parallel v_{\rm te})$, $v_{\rm te} = \sqrt{T_e/m_e}$, and
$Z(\xi)=\pi^{-1/2}\int dt 
\exp(-t^2)/(t-\xi)$ [or the analytic continuation of this for ${\rm Im}(\xi)
\le 0$] is the plasma dispersion function.
In the limit $\xi \gg 1$, the solution to the dispersion relation is a wave with frequency
\begin{align}
\omega_H &= \frac{k_\parallel v_{\rm te}}{|k_\perp| \rho_s}\label{eqn:omegaH}.
\end{align}
For $k_\perp \rho_s \ll 1$, this is a very high-frequency wave that must be handled
carefully to remain numerically stable. Note that this wave does not affect parallel
transport in the SOL because the main heat pulse propagates at the ion sound speed,
and this wave is even faster than the electrons for $k_\perp \rho_s \ll 1$.

This wave is the electrostatic limit of the shear Alfv\'en 
wave,\cite{Lee1987,Belli05} which lies in the regime of inertial Alfv\'en waves.\cite{Lysak1996,Vincena2004}
The difficulties introduced by such a wave could be eased by including
magnetic perturbations from $A_\parallel$, in which case the dispersion relation (in the
fluid electron regime $\xi \gg 1$) becomes\cite{Belli05} $\omega^2 = k_\parallel^2
v_{\rm te}^2 / (\hat{\beta}_e + k_\perp^2 \rho_s^2)$, where $\hat{\beta}_e = 
(\beta_e/2) (m_i/m_e)$ and $\beta_e = 2 \mu_0 n_e T_e / B^2$.
In the electrostatic limit $\hat{\beta}_e = 0$, we
recover \eqr{\ref{eqn:omegaH}}, but retaining a finite $\hat{\beta}_e$ would set a
maximum frequency at low $k_\perp$ of $\omega = k_\parallel v_{\rm te} / 
\hat{\beta}_e^{1/2} = k_\parallel v_A$, where $v_A$ is the Alfv\'en velocity, 
avoiding the $k_\perp \rho_s \rightarrow 0$ singularity of the electrostatic case.
(We shall defer further discussion of magnetic fluctuations to a future
paper, as that brings up another set of interesting numerical
subtleties.)

For electrostatic simulations, a modified ion polarization term can be
introduced to effectively set a minimum value for the perpendicular wave
number $k_\perp$.
This modification can be used to slow down the electrostatic shear Alfv\'en wave to
make it more numerically tractable.
(Even when magnetic fluctuations are included, one still might
want to consider an option of introducing a long wavelength modification
for numerical convenience or efficiency.)

When choosing how to select the minimum value for $k_\perp \rho_s$, it is
useful to consider the set of $k_\perp$'s represented on the
grid for particular simulation parameters.
Consider an axisymmetric system (as in Fig. \ref{fig:geoC}) with
constant $B/B_\xi$, where $B$ is the total magnetic field, and
$B_\xi$ and $B_\zeta$ are the components of $\boldsymbol{B}$ in the poloidal and
toroidal directions.
It follows that $\partial_\perp = (B_\zeta/B_\xi) \partial_\parallel$, so
\begin{align}
k_{\perp,\mathrm{max}} &= \frac{B_\zeta}{B_\xi} k_{\parallel,\mathrm{max}}.
\end{align}
The maximum parallel wavenumber can be estimated as
$k_{\parallel,\mathrm{max}} \Delta z \sim \pi N_{nc}$, 
where $\Delta z$ is the width of a single cell in position space, and
$N_{nc}$ is the total degrees of freedom per cell used in the
finite element DG representation of the position coordinate.

Therefore, one has
\begin{align}
k_{\perp,\mathrm{max}} &= \frac{B_\zeta}{B_\xi} \frac{\pi N_{nc}}{\Delta z}.
\end{align}
In our simulations, $N_{nc}=3$ and $\Delta z = 10$ m using $8$ cells in the
spatial direction to represent an $80$ m parallel length.
Assuming that $B_\xi/B = \sin(6^\circ)$, 
one estimates that $k_{\perp,\mathrm{max}}\rho_s \approx 2.5 \times
10^{-2}$ for 1.5 keV deuterium ions with $B=2$ T. Thus, the
perpendicular wave wavenumbers represented by a typical grid are fairly
small.

The general modified gyrokinetic Poisson equation we consider is of
the form
\begin{align}
-\partial_\perp ( C_\epsilon \epsilon_\perp \partial_\perp \phi)
 + s_\perp(z,t)
 (\phi - \langle \phi \rangle) = \frac{\sigma_g(z)}{\epsilon_0},
\label{eqn:mod-gk-Poisson}
\end{align}
where $s_\perp(z,t)=k^2_{\rm min}(z) {\epsilon_\perp(z,t)}$ is a
shielding factor (we allow $k_{\rm min}$ to depend on position but not
on time in order to preserve energy conservation, as described below)
and $\langle \phi \rangle$ is a
dielectric-weighted flux-surface-averaged 
potential defined as 
\begin{align}
\langle \phi \rangle &= \frac{\int dz \, s_\perp \phi}{\int dz \, s_\perp} \label{eqn:phi-avg}.
\end{align}
The fixed coefficient $C_\epsilon$ is for generality, making it easier to consider various limits later.

The sound gyroradius is chosen to be defined by $\rho_{s}^2(z,t) = c_s^2(z,t)/ \Omega_{ci}^2 =
T_e(z,t)/(m_i \Omega_{ci}^2)$, using the mass and cyclotron frequency of
a main ion species. A time-independent sound gyroradius (using a
typical or initial value for the electron temperature $T_{e0}$) is defined by
 $\rho_{s0}^2(z) = c_{s0}^2(z)/ \Omega_{ci}^2 = T_{e0}(z)/(m_i \Omega_{ci}^2)$. Note
that the shielding factor can also be written as $s_\perp(z,t)=[k_{\rm
min}(z) \rho_{s0}(z)]^2 {\epsilon_\perp(z,t)} / \rho^2_{s0}(z)$.

For simplicity, $k_{\rm min} \rho_{s0}$ is chosen to be a constant independent of
position. 
Its value should be small enough that the wave in \eqr{\ref{eqn:omegaH}} is high enough in
frequency that it does not interact with other dynamics of
interest, but not so high in frequency that it forces the explicit time
step to be excessively small. For some of our simulations, we use
$k_{\rm min} \rho_s = 0.2$,
which leads to only a 2\% correction to the ion acoustic 
wave frequency $\omega = k_\parallel c_s / \sqrt{1 + k_\perp^2 \rho_s^2}$ at
long wavelengths. Convergence can be checked by taking the
limit $k_{\rm min} \rho_{s0} \rightarrow 0$.

As a simple limit, one can even set $C_\epsilon = 0$ and keep just the $s_\perp$ term, which replaces the
usual differential gyrokinetic Poisson equation with a simpler algebraic model.
This approach should work fairly well for low frequency dynamics.
The basic idea is that for long-wavelength ion-acoustic
dynamics, the left-hand side of \eqr{\ref{eqn:mod-gk-Poisson}}
is small, so the potential is primarily determined by 
the requirement that it adjust to keep the electron density on the
right-hand side almost equal to the ion guiding center density.
(At low frequencies, the electron density is close to a Boltzmann response,
which depends on the potential.) In future work, one could consider
using an implicit method, perhaps using the method here as a 
preconditioner. Alternatively, electromagnetic effects will slow down
the high-frequency wave so that explicit methods may be sufficient.

The flux-surface-averaged potential $\langle \phi \rangle$ is subtracted
off in \eqr{\ref{eqn:mod-gk-Poisson}} so that the model polarization term is gauge invariant like the
usual polarization term.
This choice is also related to our form of the logical sheath boundary
condition, which assumes that the electron and ion guiding center fluxes
to the wall are the same so that the net guiding center charge vanishes, $\int dz \, \sigma_g = 0$.
Just as the net guiding center charge vanishes, our model polarization charge
density, $s_\perp (\phi - \langle \phi \rangle)$, also averages to zero. 
This approach neglects ion polarization losses to the wall, which is consistent in this
model because integrating \eqr{\ref{eqn:gk-Poisson}} over all space then
gives $\partial_\perp \phi = 0$ at the plasma edge.
(One could consider future modifications to account for polarization drift
losses to the wall, but the present model is found to agree fairly well
with full-orbit PIC results.)

With this approach, it is also necessary to modify the Hamiltonian in order to preserve
energy consistency with this modified gyrokinetic Poisson equation.
The modified Hamiltonian is written in the form
\begin{align}
H_s = \frac{1}{2} m_s v_\parallel^2 + q_s (\phi - \langle \phi \rangle) - \frac{1}{2} m_s \hat{V}_E^2,
\end{align}
where $\hat{V}_E^2$ is a modified $\boldsymbol{E}
\times \boldsymbol{B}$ velocity that is
chosen to conserve energy. 
The constant $\langle \phi \rangle$ term in $H_s$ has no effect on the
gyrokinetic equation because only gradients of $\phi$ matter, but it
simplifies the energy conservation calculation.
The total energy is still $W_{\rm tot} = \int dz \sum_s \int
dv_\parallel \, f_s H_s$, and its time derivative (neglecting boundary
terms that are straightforward to evaluate) can be written as
\begin{align}
\frac{d W_{\rm tot}}{dt} &= \int dz \sum_s \int dv_\parallel \, f_s \frac{\partial H}{\partial t} \nonumber \\
 &= \int dz \left( \sigma_g \frac{\partial}{\partial t} \left( \phi -
\langle \phi \rangle \right) 
 - \sum_s \frac{1}{2} n_s m_s \frac{\partial}{\partial t} \hat{V}_E^2 \right).
\end{align}
Using the modified gyrokinetic Poisson equation (\ref{eqn:mod-gk-Poisson}) and
integrating the first term by parts gives
\begin{align}
\frac{d W_{\rm tot}}{dt} 
 &= \int dz \Bigg( \sum_s \frac{1}{2} n_s m_s
C_\epsilon \frac{\partial}{\partial t} V_E^2 \nonumber\\
 & +\frac{\epsilon_0}{2} s_\perp \frac{\partial}{\partial t} 
 (\phi - \langle \phi \rangle)^2 - \sum_s \frac{1}{2} n_s m_s \frac{\partial}{\partial t} \hat{V}_E^2 \Bigg),
\end{align}
so energy is conserved if one chooses
\begin{align}
\hat{V}_E^2 &= C_\epsilon V_E^2 + \frac{\epsilon_0 s_\perp}{\sum_s n_s m_s} (\phi - \langle
\phi \rangle)^2
\end{align}
and require that the coefficient $\epsilon_0 s_\perp/ (\sum_s n_s m_s)$ be
independent of time so that it comes outside of a time derivative.
Using \eqr{\ref{eqn:epsilon_perp}} and the definition of $s_\perp$ after
\eqr{\ref{eqn:mod-gk-Poisson}}, one sees that $\epsilon_0 s_\perp/ (\sum_s n_s m_s) =
k_{\rm min}^2(z) /B^2$, which is indeed independent of time because $k_{\rm
min}$ was chosen not to have any time dependence.

In the limit that one uses only the algebraic model polarization term with $C_\epsilon=0$, one finds that
\begin{align}
\hat{V}_E^2 = (k_{\rm min} \rho_{s0})^2 \left( \frac{e \delta \phi}{T_{e0}} \right)^2 c_{s0}^2,
\end{align}
where $\delta \phi = \phi - \langle \phi \rangle$.
For $k_{\rm min} \rho_{s0} = 0.2$ and $e \delta \phi / T_{e0} \sim 1$, this $\boldsymbol{E}
\times \boldsymbol{B}$ energy
could be order 4\% of the total energy.

\section{Numerical implementation details\label{sec:numerical}}
One detail of solving the modified gyrokinetic Poisson equation (\ref{eqn:mod-gk-Poisson}) is how to
determine the flux-surface-averaged component, which is related to the boundary conditions.
Consider the case in which $\epsilon_\perp = 0$, and expand $\phi =
\langle \phi \rangle + \delta \phi$. Then $\delta \phi$ is determined by
the algebraic equation
\begin{align}
s_\perp(z) \, \delta \phi(z) = \frac{\sigma_g(z)}{\epsilon_0}\label{eqn:algebraicPoisson}.
\end{align}
Imposing the boundary condition that the value of $\phi$ at the plasma edge be
equal to the sheath potential gives $\phi(z_R) = \phi_s = \langle \phi \rangle + \delta \phi(z_R)$
(the left and right boundaries have been assumed to be symmetric here), which gives an additional equation
to determine $\langle \phi \rangle$.
The final expression is 
\begin{align}
\phi(z) = \delta \phi(z) - \delta \phi(z_R) + \phi_s.
\end{align}

In order to maintain energy conservation, it is important that the algorithm preserve the numerical
equivalent of certain steps in the analytic derivation.
In our algorithm, based on Liu and Shu's \cite{LiuShu2000} algorithm for
the incompressible Euler equation, $\phi$ must be obtained using
continuous finite elements, although the charge density $\sigma_g$ is discontinuous in our Poisson equation.

To preserve the integrations involved in energy conservation, it is important
to ensure that one can multiply \eqr{\ref{eqn:algebraicPoisson}} by the fluctuating potential,
integrate over all space, and preserve
\begin{align}
\int dz \, \delta \phi \, s_\perp \, \delta \phi = \frac{1}{\epsilon_0} \int dz \, \delta \phi \, \sigma_g.
\end{align}
This requirement ensures that a potential part of the energy on the right-hand side is exactly related to a
field-like-energy on the left-hand side.
This quantity will be preserved if one projects the modified Poisson equation onto all of the continuous basis functions
$\psi_j$ that are used for $\phi$ (i.e., $\phi(z) = \sum_j \phi_j
\psi_j(z)$) to ensure that
\begin{align}
\langle \psi_j s_\perp \phi \rangle = \langle \psi_j \sigma_g \rangle.
\end{align}

For piecewise linear basis functions, this leads to a tri-diagonal equation for $\phi_j$
that has to be inverted to determine $\phi$. Because $s_\perp \propto n(z,t)$ varies in time,
this will take a little bit of work, but as one goes to higher dimensions
in velocity space, the Poisson solve (which is only in the
lower-dimensional configuration space)
will be a negligible fraction of the computational time.

\subsection{Boundary Conditions}
Gyrokinetics does not need to resolve the restrictive 
Debye length ($\sim\lambda_{De}$) or plasma frequency time scales
($\sim\omega_{pe}^{-1}$), so the sheath is usually not directly
resolved. Instead, the effects of the sheath can be incorporated through
the use of logical sheath boundary conditions.\cite{Parker1993} 
For a normal positive sheath, all incident ions flow into the wall, but
incident electrons with energies below the sheath potential are reflected
back into the domain such that 
there is zero net current into the wall.
(For biased endplates or higher dimensional problems with non-insulating
walls, one could consider more general boundary conditions that involve
currents in and out of the wall at various places.)
At the right boundary, for example, this condition is expressed as
\begin{align}
\int_0^\infty dv\thinspace v f_i(z_R,v,t) & = \int_{v_c}^\infty dv\thinspace v f_e(z_R,v,t),
\end{align}
where $z_R$ is the coordinate of the domain edge.
The cutoff velocity $v_c>0$ is determined numerically through a search algorithm.
The sheath potential is then determined using the relation $e\phi_s = m_e v_c^2/2$.

In order to reflect all electrons incident on the sheath with velocity in the range
$0 < v < v_c$, the electron distribution function in this range is copied into
ghost cells according to
\begin{align}
f_e(z_R,-v,t) &= f_e(z_R,v,t), \quad 0<v<v_c,
\end{align}
and $f_e(z_R,-v,t) = 0$ for $v>v_c$. This condition can also be written as
$f_e(z_R,-v,t) = f_e(z_R,v,t) H(v_c -v)$ for $v > 0$.
This condition results in the reflection of electrons with velocity in the range
$0 < v < v_c$ back into the domain with the opposite velocity, while
the electrons with energy sufficient to overcome the sheath potential
will flow out of the system to the divertor plates.

The implementation of logical sheath boundary conditions needs a slight modification
for use in a continuum code.
Typically, the cutoff velocity will fall within a cell and not exactly on a cell edge.
A direct projection of the discontinuous reflected distribution onto
the basis functions used in a cell could lead to negative values of the
distribution function at some velocities in the cell. Future work could
consider methods of doing higher-order projections that incorporate
positivity constraints, but for now we have used a simple scaling
method, in which the entire distribution function inside the ``cutoff
cell'' is copied into the ghost cell and then scaled by the fraction
required to ensure that the electron flux at the domain edge equals the ion flux.
For scaling the reflected distribution function in the cutoff cell on
the right boundary, this fraction is
\begin{align}
c &= \frac{\int_{v_j - \Delta v/2}^{v_c}dv\, v f_e(z_R,v,t)}{\int_{v_j-\Delta v/2}^{v_j+\Delta v/2}dv \, vf_e(z_R,v,t)},
\end{align}
where $\Delta v$ is the cell width in velocity space, and $v_j$ denotes the center of the cell.

\section{Simulation Results\label{sec:results}}
The main parameters used for our simulations were described in
\refcite{havlickova} and were chosen to model an ELM on the JET
tokamak for a case in which the density and temperature at the top of the
pedestal were $n_{\rm ped} = 5 \times 10^{19}$ m$^{-3}$ and $T_{\rm ped} =
1.5$ keV. The ELM is modelled as an intense particle and heat source in
the SOL that lasts for 200 $\mu$s, spread over a poloidal length of 2.6
m around the midplane (as described below) and a radial width in the
SOL of 10 cm. The model SOL has a major radius of 3 m, and this source
corresponds to a total ELM energy of about 0.4 MJ. The simulation
domain has a length of $2 L_{||} = 80$ m, the length of a magnetic field
line in the SOL, with a field line pitch of $6^\circ$.
The kinetic equation with the source term on the right-hand side is
\begin{align}
\frac{\partial f}{\partial t} -\{H,f\}
= g(t) \, S(z) \, F_M\boldsymbol{(}v_\parallel, T_{S}(t)\boldsymbol{)}\label{eqn:sim},
\end{align}
where $F_M\boldsymbol{(}v_\parallel, T_S(t)\boldsymbol{)}$ is a unit Gaussian in variable $v_\parallel$
with a time-dependent temperature $T_S(t)$.
The function $S(z)$ is the same for both particle species, and is represented as
\begin{align}
S(z) &= \begin{cases} S_0 \cos\left(\frac{\pi z}{L_s}\right) & |z| < \frac{L_s}{2}\\ 0 & \mathrm{else} \end{cases},
\end{align}
where $L_s = 25$ m is length of the source along the magnetic field line.
The value of $S_0$ was computed using the scaling\cite{havlickova}
\begin{align}
S_0 = A \, n_{ped} \, c_{s,ped} / L_s,
\end{align}
where the constant of proportionality $A$ was chosen to be $1.2 \sqrt{2} \approx 1.7$ for comparison with \refcite{havlickova}.
In our simulations, $S_0 \approx 9.066\times 10^{23}$ m$^{-3}$\thinspace s$^{-1}$.
\begin{figure}
\includegraphics[width=0.45\textwidth]{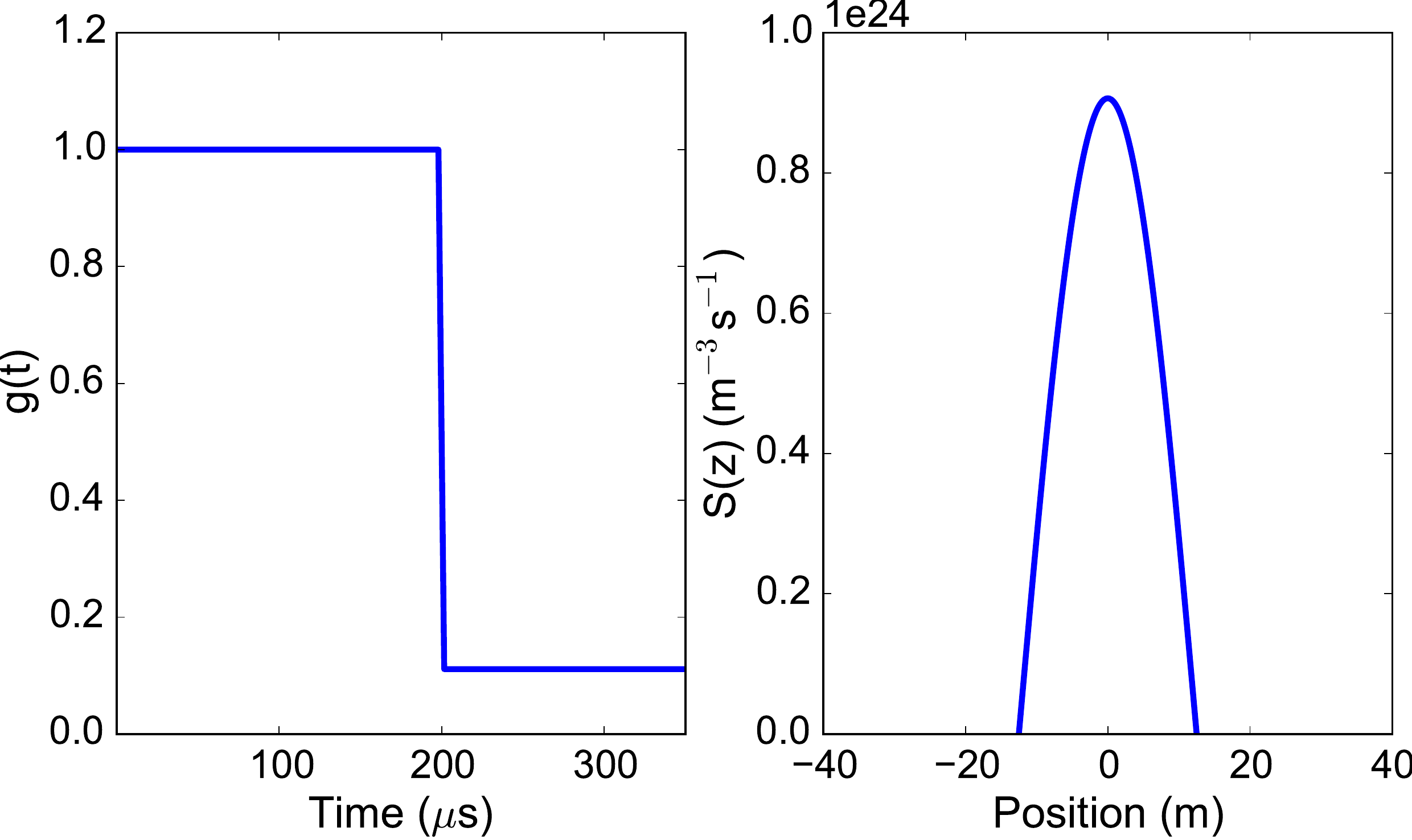}
\caption{\label{fig:source} Spatial and temporal profiles of the source term on the right-hand side
of \eqr{\ref{eqn:sim}}.}
\end{figure}

The function $g(t)$ in \eqr{\ref{eqn:sim}} is used to model the time-dependence of the particle source:
\begin{align}
g(t) &= \begin{cases} 1 & 0 < t < 200 \, \mu\mathrm{s}\\ 1/9 & t > 200 \, \mu\mathrm{s} \end{cases}.
\end{align}

The post-ELM source also has reduced electron and ion temperature, represented
by the $T_S(t)$ parameter in the Maxwellian term $F_M$ in \eqr{\ref{eqn:sim}}, which
has the value $1.5$ keV from $0 < t < 200$ $\mu$s for both ions and electrons.
The electron temperature for $t > 200$ $\mu$s is 210 eV, and the ion temperature
is reduced to 260 eV. The end time for the simulation is $t=350$ $\mu$s.

We performed our simulations using second-order serendipity basis
functions \cite{Arnold2011} on a grid with 8 cells in the spatial direction and 32
cells in the velocity direction. (In 1D, second-order basis functions correspond
to piecewise parabolic basis functions, or 3 degrees of freedom within
each cell.) The case with kinetic electrons and ions takes only about three
minutes to run on a standard laptop, although we have not yet extensively optimized
our code.

\subsection{Initial Conditions}
In previous papers that looked at this problem, the codes were typically
run for a while with the same weak source that would be used in the
post-ELM phase to reach a quasi-steady state before the intense ELM
source was turned on. The authors found that the final results were not very
sensitive to the duration of the pre-ELM phase or the initial conditions
used for it. However, there is formally no normal steady state for this
problem in the collisionless limit (low energy particles build up over time
without collisions). To remove a possible source of ambiguity for
future benchmarking, here we specify more precise initial
conditions chosen to approximately match initial conditions at the
beginning of the ELM phase used in previous work.

We model the initial electron distribution function as
\begin{align}
f_{e0}(z,v_\parallel) &= n_{e0}(z) F_M(v_\parallel, T_{e0}),
\end{align}
with $T_{e0} = 75$ eV.
The electron density profile (in $10^{19}$ m$^{-3}$) is defined as
\begin{align}
n_{e0}(z) &= 0.7 + 0.3\left(1-\left|\frac{z}{L_\parallel}\right|\right)\nonumber\\
& \qquad{}+0.5\cos\left(\frac{\pi z}{L_s}\right)H\left(\frac{L_s}{2} - |z|\right).
\end{align}

The initial ion distribution function is modeled as
\begin{equation}
f_{i0}(z,v_\parallel) = \begin{cases} F_L & z < -\frac{L_s}{2}\\
\begin{aligned}
&\Bigg[\left(\frac{1}{2}- \frac{z}{L_s}\right) F_L \\
&\quad{}+ \left(\frac{1}{2} + \frac{z}{L_s}\right) F_R\Bigg]
\end{aligned} 
& -\frac{L_s}{2} < z < \frac{L_s}{2}\\
F_R& z > \frac{L_s}{2} \end{cases}.
\end{equation}
Here, $F_L$ and $F_R$ are left and right half-Maxwellians defined as
\begin{align}
F_R(z,v_\parallel; T_{i0}) &= \hat{n}(z)F_M(v_\parallel, T_{i0})H(v_\parallel),\\
F_L(z,v_\parallel; T_{i0}) &= \hat{n}(z)F_M(v_\parallel, T_{i0})H(-v_\parallel),
\end{align}
where $\hat{n}(z) = 2n_{i0}(z)$, $H$ is the Heaviside step function, and
the initial ion temperature profile (in eV) is defined as
\begin{align}
T_{i0}(z) &= 100 + 45\left(1-\left|\frac{z}{L_\parallel}\right|\right)\nonumber\\
& \qquad{}+30\cos\left(\frac{\pi z}{L_s}\right)H\left(\frac{L_s}{2} - |z|\right).
\end{align}

The expressions for the $n_{e0}$ and $T_{i0}$ profiles were chosen to
approximate those described in private communication with the author of
\refcite{havlickova},\cite{havlickovaEmail} which were originally
obtained from simulations that had run for a while with a weaker source
to achieve a quasi-steady state before the strong ELM source was turned
on, as described at the beginning of this subsection.

Given an initial electron density profile, we then calculate an initial
ion guiding center density profile to minimize the excitation of
high-frequency kinetic Alfv\'{e}n waves.
We do this by choosing the initial ion guiding-center density $n_i(z)$
so that it gives a potential $\phi(z)$ that results in the electron
density's being consistent with a Boltzmann equilibrium, i.e., the
electrons are initially in parallel force balance and do not excite
high-frequency kinetic Alfv\'en waves.
A Boltzmann electron response is
\begin{align}
n_e(z) &= C \exp \left(\frac{e \phi(z)}{T_e}\right).
\end{align}

Taking the log of the above equation and then an $n_e$-weighted average, one has
\begin{align}
\langle \log n_e \rangle_{n_e} &= \log C + \frac{e\langle\phi\rangle_{n_e}}{T_{e0}},
\end{align}
where $T_e$ has been assumed to be a constant $T_{e0}$.

Note that one is free to add an arbitrary constant to $\phi$ since only
gradients of $\phi$ affect the dynamics.
Choosing the additional constraint that $\langle \phi \rangle_{n_e} =
0$, one can express the constant $C$ in terms of $n_e$. (This convention
for $\langle \phi \rangle_{n_e}$ is only for convenience, as any
constant can be added to $\phi$ in the plasma interior without affecting
the results. After the first time step, the sheath boundary condition
will be imposed, which will give a non-zero value for the average
potential.)

One then has the following equation for $\phi$:
\begin{align}
\frac{e\phi}{T_{e0}} &= \log n_e - \langle \log n_e \rangle_{n_e}.
\end{align}

This $\phi$ can be used with the gyrokinetic Poisson equation to solve for $n_i(z)$ by iteration.
With a small $m_e/m_i$ ratio, the gyrokinetic Possion equation can be written as
\begin{align}
n_i(z) \left(1 - k_\perp^2 \rho_{s0}^2 \frac{e(\phi - \langle \phi \rangle_{n_i})}{T_{e0}}\right) &= n_e(z),
\end{align}
where with the small $m_e/m_i$ ratio approximation, the dielectric-weighted
average is equivalent to an ion density-weighted average.
The left-hand side of this equation is a nonlinear function of $n_i$
(because it appears as a leading coefficient and in the density-weighted
average $\langle \phi \rangle_{n_i}$), which is solved for by using iteration:
\begin{align}
n_i^{j+1}(z) &= \frac{n_e(z)}{1 - k_\perp^2 \rho_{s0}^2 \frac{e}{T_{e0}}
 \left(\phi - \langle \phi \rangle_{n_i^{j}} \right)}.
\end{align}

Note that the the averaged $\phi$ on the right-hand side is weighted by $n_i^j$,
the previous iteration's ion density.
Convergence can be improved by adding a constant to $n_i(z)$ each
iteration to enforce global neutrality $\langle n_i \rangle = \langle
n_e \rangle$.
In our tests, the initial ion density profile was calculated to $10^{-15}$ relative error in five iterations.

\subsection{Divertor heat flux with drift-kinetic electrons}
\begin{figure}
\includegraphics[width=0.45\textwidth]{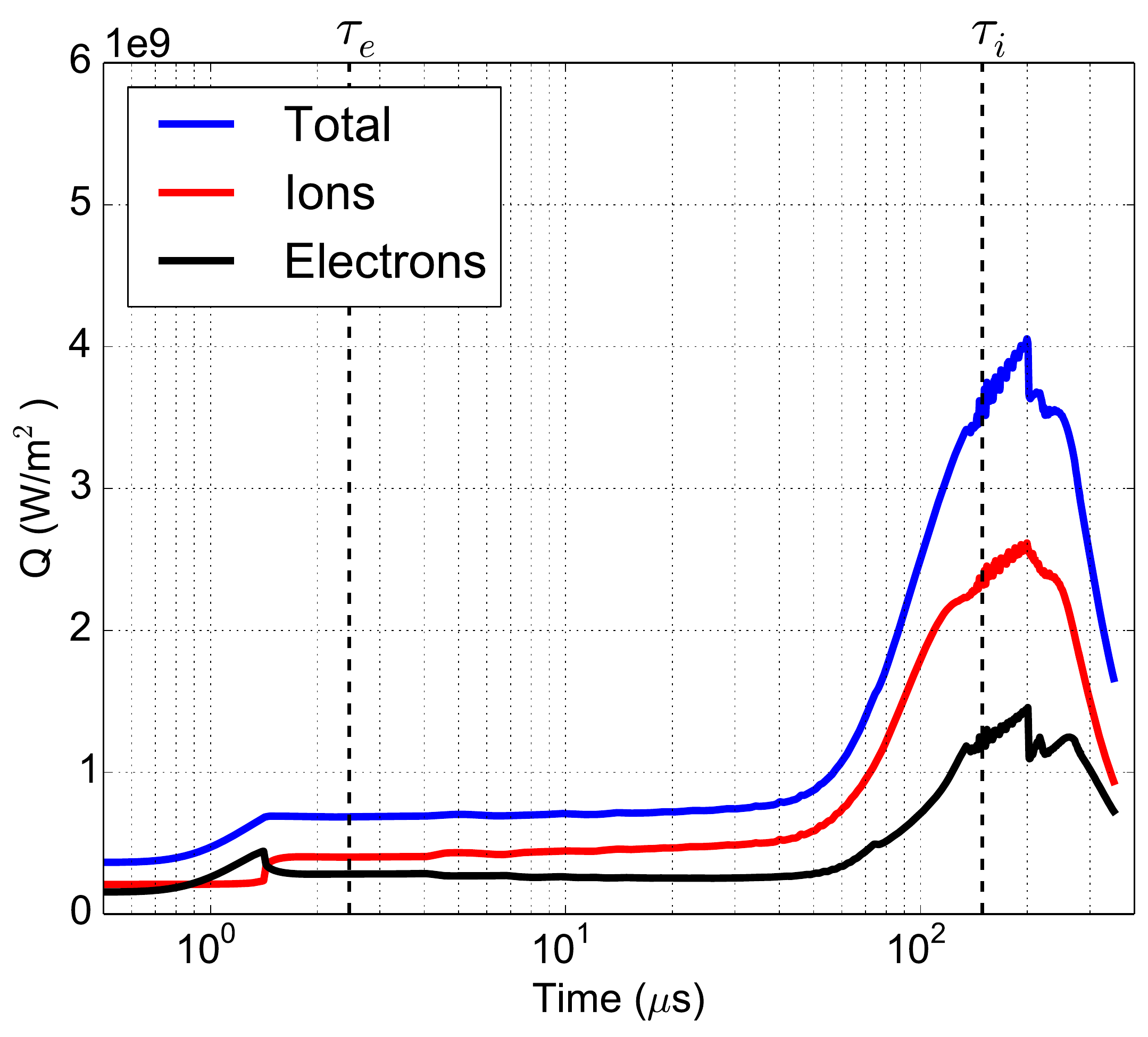}
\caption{\label{fig:heat-flux-es}Parallel heat flux at the divertor
 plate vs. time with drift-kinetic electrons. The electron and ion
 thermal transit times $\tau_e$ and $\tau_i$ are indicated by the
 vertical dashed lines.}
\end{figure}
Figure \ref{fig:heat-flux-es} shows the parallel heat flux on the target plate vs. time
using the 1D electrostatic model with a fixed $k_\perp \rho_{s0}=0.2$.
A rapid response in the electron heat flux is observed at early times,
on the order of the electron transit timescale $\tau_e \sim
L_\parallel/v_{\rm te, ped} \approx 2.46$ $\mu$s.
This response is due to fast electrons reaching the target plate, 
which initially cause a modest rise in the electron heat flux from $t \sim 1$
$\mu$s to $t \sim 1.5$ $\mu$s.
This build-up of fast electrons result in a rise in the sheath potential at $t \sim 1.5$ $\mu$s, which
causes a modest rise in the ion heat flux
and a modest drop in the electron heat flux until the arrival of the
bulk ion heat flux at a later time.
We did a scan in $k_\perp^2 \rho_{s0}^2$ over a factor of 20 (from $k_\perp^2 \rho_{s0}^2 = 0.04$ to 0.1) and found only a few percent variation in the resulting plot of heat flux vs. time, verifying that the results are not sensitive to the exact value of this parameter (as long as it is small).


As pointed out in a recent invited talk,\cite{Leonard2014} one of the
original motivations for calculations of this kind (such as
\refcite{Pitts2007}) was a concern that the fast parallel thermal transport of
electrons would cause a very large heat flux to arrive at the divertor
plates on the electron transit time scale. Our results confirm the
previous calculations that found that although there is a modest rise in
the heat flux on the electron transit time scale, the sheath
potential (and the potential variation along the field line) increases
to confine most of the electrons so that the bulk of the ELM energy arrives
at the target plate only on the slower ion time scale. (Nevertheless,
even this ELM power is so large that erosion of solid target plates is a
concern, and methods of mitigating or avoiding ELMs are being studied.)

The bulk of the ELM energy is carried by the ions, which arrive at the
target plate on the order of the ion thermal transit timescale,
$\tau_i \sim L_\parallel/v_{\rm ti} \approx 149$ $\mu$s.
The reduction of source strength and temperature after $200$ $\mu$s results
in the abrupt drop seen in the electron heat flux.

The parallel heat flux (parallel to the magnetic field) on the right
target plate for each species is calculated as
\begin{align}
Q_s &= \frac{1}{2}m_s \int_{v_{c,s}}^{\infty} dv \, f_s \, v^3 + (T_\perp + q_s\phi_s) \int_{v_{c,s}}^{\infty} dv \, f_s \, v,
\end{align}
where $v_{c,s} = \sqrt{\mathrm{max}(-2 q_s \phi_s/m_s, 0)}$ accounts
for the reflection of electrons by the sheath.
The $q_s \phi_s$ term in the second integral models the acceleration
of ions and deceleration of electrons as they pass through the sheath
to the divertor plate, a region that is not resolved in our models.
We have assumed that each species has a constant perpendicular
temperature $T_\perp = T_{ped}$ for comparison with the 1D Vlasov
results in \refcite{havlickova}.
Note that the pitch angle of the magnetic field is not factored into
this measure of heat flux on the target plate.
The heat flux normal to the target plate is $Q_{s,n} = Q_s\sin(\theta)$,
where $\theta$ is the (usually very small) angle between the magnetic field
and the surface.

Figure \ref{fig:heat-flux-es} agrees well with the $1x$+$1v$ Vlasov and full
$1x$+$3v$ PIC results in \refcite{havlickova}, providing a useful benchmark
for these codes and supporting the accuracy of the sheath boundary
conditions and the gyrokinetic-based model used here. (The small differences
between our $1x$+$1v$ results, the Vlasov results, and the PIC results
are probably due to small differences in initial conditions and the
inclusion of collisions in the PIC code.)

\subsection{Divertor heat flux with Boltzmann electron model}
We have also investigated a model that includes the effect of kinetic ions
but assumes a Boltzmann response for the electrons.
Specifically, the electron density takes the form
\begin{align}
n_e(z) &= n_{e}(z_R) \exp \left(\frac{e(\phi-\phi_s)}{T_e}\right),
\end{align}
where $n_{e}(z_R)$ is the electron density evaluated at the domain edge.
This expression can be inverted to give another algebraic equation to determine
the potential, similar to the electrostatic gyrokinetic model with a fixed $k_\perp \rho_{s0}$.
Since the time step is set by the ions, these simulations have an execution time a factor of
$\sim \sqrt{m_i/m_e}$ faster than the gyrokinetic simulation.
This property makes the Boltzmann electron model useful as a test case for code development and debugging.

The sheath potential $\phi_s$ can be determined by assuming that $f_e$ at the
target plate is a Maxwellian with temperature $T_e$.
By using logical sheath boundary conditions and quasineutrality,
\begin{align}
\phi_s &= -\frac{T_e}{e} \log \left(\frac{\sqrt{2\pi}\Gamma_i}{n_i
 v_{\rm te}}\right),
\end{align}
where $\Gamma_i$ is the outward ion flux, and all quantities are evaluated at the domain edge.
For simplicity, we selected $T_e$ in our simulations to be the
field-line-averaged value of the ion temperature $T_i(z)$, but more accurate
models for $T_e$ could be used. 

\begin{figure}
\includegraphics[width=0.45\textwidth]{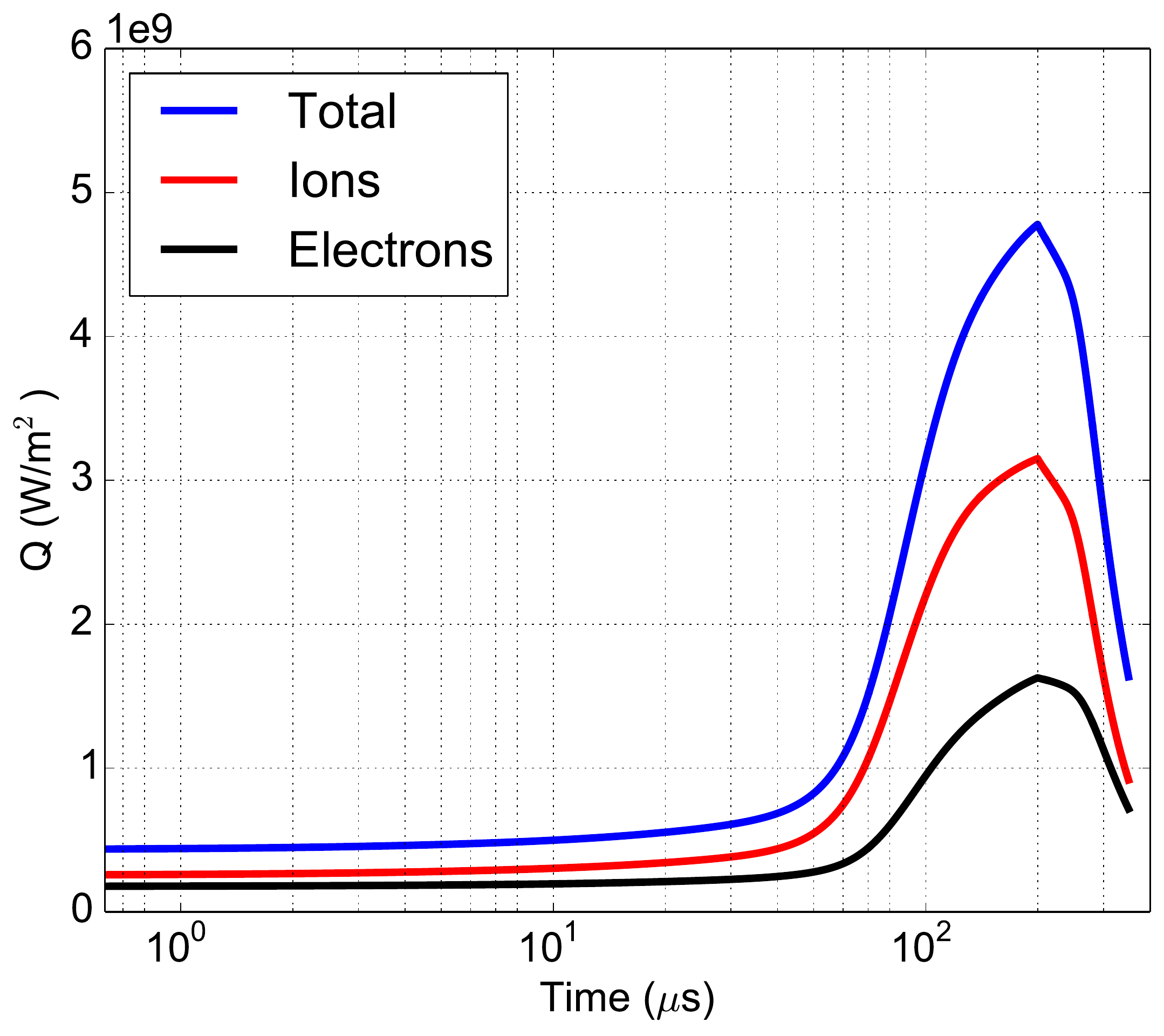}
\caption{\label{fig:heat-flux-be}Parallel heat flux at the divertor
 plate vs. time from the Boltzmann electron model.} 
\end{figure}

Figure \ref{fig:heat-flux-be} shows the parallel heat flux on the target
plate vs. time using Boltzmann electrons. As expected, kinetic electron effects
present in Fig. \ref{fig:heat-flux-es} are not resolved by this model.
When compared to a simulation using kinetic electrons, the main heat flux
at $t \sim 100-200$ $\mu$s is predicted fairly well by the Boltzmann electron model.

The expression for the electron parallel heat flux on the target plate is calculated as
\begin{align}
Q_e &= \frac{1}{2}m_e \int_{v_c}^{\infty} dv \, f_e \, v^3 + (T_\perp - e\phi_s) \int_{v_c}^{\infty} dv \, f_e \, v \nonumber\\
&= (T_e + T_\perp) \int_{0}^{\infty} dv \, f_i \, v.
\end{align}

\section{Conclusions\label{sec:conclusions}}
We have used a gyrokinetic-based model to simulate the propagation
of a heat pulse along a scrape-off layer to a divertor target plate.
We have described a modification to the ion polarization term to slow
down the electrostatic shear Alfv\'{e}n wave.

Our main results include the demonstration that this gyrokinetic-based
model with logical sheath boundary conditions is able to agree well with
Vlasov and full-orbit (non-gyrokinetic) PIC simulations, without needing
to resolve the Debye length or plasma frequency. This simplification allows the
spatial resolution to be several orders of magnitude coarser than the
electron Debye length (and the time step several orders of magnitude
larger than the plasma period) and thus leads to a much faster
calculation. Our results also confirm previous work that the
electrostatic potential in this problem varies to confine most of the
electrons on the same time scale as the ions, so the main ELM heat
deposition occurs on the slower ion transit time scale.

Additionally, we have described a model using Boltzmann electrons
that is useful for code development and debugging.
This model does not include kinetic electron effects but runs
much faster than simulations with kinetic electrons and ions.

Although this paper focuses on electrostatic simulations,
we have also extended our simulations to include magnetic fluctuations.
These extensions involve a number of interesting and subtle physics
and algorithm issues that will be described in a future paper.

Since we have assumed only a single $k_\perp$ mode in our simulations to
limit the high frequency of the electrostatic shear Alfv\'{e}n wave,
future work can include allowing a spectrum of $k_\perp$ modes.
For 1D electromagnetic simulations, this modification requires inverting
the $\nabla_\perp^2$ operators that appear in the gyrokinetic Poisson
equation and Ampere's law.
We defer further discussion of this to a future paper because including
a magnetic component to the fluctuations will be important when a
spectrum of very low $k_\perp$ modes is kept in order to limit on the
frequency of the shear Alfv\'{e}n wave at low $k_\perp$.

Future work on these models can also include extensions to higher spatial
and velocity dimensions. An axisymmetric 2D model can use a specified
diffusion coefficient to model radial transport in the SOL. A full 3D
gyrokinetic model would include turbulence, so radial transport can be
self-consistently calculated. These models could eventually include more
detailed effects such as collisions, recycling, secondary electron emission,
charge-exchange, and radiation, and could be used to study different types
of divertor configurations, including the possible usage of liquid metal coatings.

\begin{acknowledgments}
This work was supported by the U.S. Department of Energy through the Max-Planck/Princeton Center for Plasma Physics,
the SciDAC Center for the Study of Plasma Microturbulence,
and the Princeton Plasma Physics Laboratory under Contract No. DE-AC02-09CH11466.
\end{acknowledgments}

\bibliography{shi-sol-2014-arxiv}
\end{document}